# THAAD: Efficient Matching Queries under Temporal Abstraction for Anomaly Detection

Roni Mateless, Michael Segal and Robert Moskovitch

Ben-Gurion University of the Negev, Israel

mateless@post.bgu.ac.il, segal@bgu.ac.il, robertmo@bgu.ac.il

**Abstract**
In this paper we present a novel algorithm and efficient data structure for anomaly detection based on temporal data. Time-series data are represented by a sequence of symbolic time intervals, describing increasing and decreasing trends, in a compact way using gradient temporal abstraction technique. Then we identify unusual subsequences in the resulting sequence using dynamic data structure based on the geometric observations supporting polylogarithmic update and query times. Moreover, we introduce a new parameter to control the pairwise difference between the corresponding symbols in addition to a distance metric between the subsequences. Experimental results on a public DNS network traffic dataset show the superiority of our approach compared to the baselines.

## 1   Introduction

Anomaly detection is a technique used to identify unusual deviations from the expected behavior, called often outliers. Identifying anomalies enables to detect potentially malicious activities in data. These include detection of unusual changes to the normal levels of network traffic [1], identification of normal pattern for transactions or revenue [2], recognizing temporal anomalies in vehicle traffic data [3], disease diagnosis in medical image analysis [4] as well as other [5]. There is a variety of techniques that are used to identify anomalies using different statistical methods and machine learning based approaches including recent supervised and unsupervised solutions [1, 6-10]. Each of the approaches has its own pros and cons and is suited to anomaly detection in a specific domain under the specific scenario.

The typical approach to identify unusual subsequences of a given sequence is by using Euclidean distance as the similarity measure between the subsequences [12-15]. The major drawback of this approach for anomaly detection is that it might misinterpret differences in a few symbols. It means that for relatively long subsequences with a few different corresponding symbols, the Euclidean distance would be small while these differences indeed represent an anomaly. In order to overcome this difficulty, our goal is to define an additional parameter to control the pairwise difference between the corresponding symbols of the subsequences.

Temporal abstraction proposes to transform time point data [19] into symbolic time intervals. One of the main methods for temporal abstraction is gradient abstraction, in which according to the first derivative in each time point, adjacent symbols having the same value of I-Increasing/D-Decreasing/S-Stable are concatenated into symbolic time intervals. Gradient abstraction has a great potential to be very effective in anomaly detection since the absolute time-points values have no meaning by themselves. Fortunately, when considering a trend, we can make a decisive conclusions. Any change in trend whether it is small or big could be a part of an abnormal behavior. Furthermore, usually, the information contains a large amount of data, with most of it being irrelevant for anomalies. The focus on trend rather than on data itself may greatly simplify the task of anomaly detection.

To identify unusual subsequences in the sequence, effective algorithms and data structures are used supporting polylogarithmic update and query times. We obtain a better theoretical worst-case runtime solution to the HOT SAX [20-21] approach which looks for the most unusual subsequence in a given sequence and have worst-case quadratic running time (in terms of sequence length). Subsequences comparison can be smartly ordered for effective pruning using various methods, see [20-25]; however, in the worst-case scenario, when we are interested in producing the perfect ordering, the running time remains quadratic. Our new data structure allows us to break this quadratic worst-case runtime barrier.

Our solution is fast in every step, easy to extend to different types of data, easy to maintain since there is no model to train, generic and relevant for various domains. The contributions of this paper are the following:

1. THAAD – a new algorithm and an efficient data structure to identify anomalies in multivariate temporal data that went through temporal abstraction.
2. This data structure can be adjusted to work under different distance measures and can be used to improve related results that identify outlier subsequences in time series, e.g. [20-21].
3. We introduce additional parameter (see Definition 10 and Section 4) that allows us to control the pairwise difference between the corresponding symbols in subsequences in addition to the standard distance metric. This provides more flexibility and better precision in the derived results.
4. We support the theory by rigorous evaluation on a comprehensive DNS traffic public dataset, including comparison of THAAD to the baselines.

This paper is organized as follows. We start with the background, then we present our approach including formal definitions and the detailed algorithm explanation. In Section 4 we present the construction and analysis of efficient data structure allowing us significantly speed up the solution. Next, we evaluate the proposed solution with a number of experiments on network traffic, comparing it with a previously known relevant baselines. Finally, we conclude the paper.

## 2 Background

To make the basis for our research, we first review anomaly detection studies for temporal data. Next we discuss temporal abstraction techniques and, finally, relevant literature related to DNS anomalies identification including unsupervised methods is presented.

### 2.1 Anomaly Detection for Temporal data

In the last two decades, several studies provided an extensive overview of outlier detection techniques [26-28] and outlier detection based on temporal data [5]. Temporal outlier analysis that examines anomalies in the behavior of the data across time. In [5] the authors split the outlier analysis problems in temporal data to a few categoriess: Time series data, Data streams, Distributed data, Spatio-Temporal data and Network data. Time series data contains two topics, Time series Databases and Given time series, where the later are divided into point and subsequence outliers.

Additional studies introduced several techniques to identify outlier subsequences in a given time series [78-80]. Keogh et al. [20-21] consider all possible subsequences $s \in S$ of given sequence $S$ and compute the distance of each such $s$ with each other non-overlapping $s' \in S$. The subsequence $s$ is set as outlier if it has the largest distance among other subsequences $s' \in S$ [20]. Additionally, they suggested heuristic techniques to reduce the order of magnitude from quadradic runtime using Top-K pruning and heuristic reordering of candidate subsequences [21,29]. To compute the distance between subsequences, most methods use Euclidean distance while Compression based Dissimilarity Measure (CDM) is used as a distance measure in [30].

Other papers proposed to identify varying length anomalous subsequences. Chen et al. [31] proposed multi-scale anomaly detection algorithm based on infrequent pattern of time series. They defined that the anomaly pattern is the most infrequent time series pattern with the same slope and length, which is the lowest supported pattern. Senin et al. [32] used grammar induction to aid anomaly detection without any prior knowledge. The authors transformed time series values into symbolic form, inferred a context free grammar, and exploited its hierarchical structure to discover anomalies.

### 2.2 Temporal Abstraction

Temporal Abstraction proposes to transform time point based data [19] into symbolic time intervals, for which specific methods are being developed [33-35]. There are two main types of methods: state and gradient abstraction. In state-abstraction, the time point values are classified (discretized) into states (Low/Medium/High). These cut-offs can be based on common standards in the domain, or data driven. In gradient abstraction, they are discretized according to the first derivative (I-Increasing/D-Decreasing/S-Stable). Then, after each time point is classified into the appropriate state or gradient symbol, if adjacent symbols are the same, they are concatenated into symbolic time intervals. Obviously, the number of state/gradient symbols can be other than three depending on the required granularity and generalization. Unlike gradient abstraction, which is based on the first derivative of a sequence of values, state abstraction can be applied in different ways based on the values of the cut-offs, for which several sources are optional. The most intuitive source is knowledge-based (KB) [11], in which the cut-off values come from the common practice in the domain knowledge. Alternatively, the cut-offs can be learned from the data. The simplest method is Equal Width Discretization (EWD), in which the range of values is split equally into states. Another intuitive method that has become increasingly popular is Symbolic Aggregate approXimation (SAX) [33], in which the Gaussian distribution of the time series determines the states and is defined by the mean and standard deviations. Persist [34] is an unsupervised method that was designed to maximize the duration of the resulting time intervals, to facilitate the later discovery of coinciding time interval patterns [35]. The Temporal Discretization for Classification (TD4C) is a supervised data-driven method [36], which determines the cut-offs that mostly differentiate classes based on the states distributions of each class along time.

### 2.3 DNS Anomalies Identification

Network traffic is based on multivariate time-series data, that describe packets flow over time. Common DNS attacks that are known to the community are: (D)DoS – when one or more attackers controlling one or more devices launch an avalanche of messages to one or more DNS servers; Cache poisoning – when a query is sent to a local DNS server and local DNS returns the fake response to the resolver and caches the forged mapping; Tunneling – when the DNS packets can be used to create a hidden data channel (covert channel); Fast-flux – when someone hides critical hosts behind a changing set of compromised hosts; Zone transfer hijacking – when the attacker just pretends that he is a slave and asks the master for a copy of the zone records; Dynamic update corruption – when a non-authorized machine can update the DNS record.

Several studies dealt with the identification of DNS anomalies. As mentioned above, popular techniques used by cyber-criminals to hide their critical systems is fast-flux. The ICANN Security and Stability Advisory Committee [37] released a paper giving a clear explanation of the technique. Nazario and Holz [16] performed some interesting measurements on known fast-flux domains. Villamarn-Salomn and Brustoloni [38] focused their detection on abnormally high or temporally concentrated query rates of dynamic DNS queries. The research by Choi et al. [39] created an algorithm that checks multiple botnet characteristics. The detection is based on Dynamic DNS, fixed group activity and a mechanism for detecting migrating C&C servers. Born and Gustafson [40] researched a method for detecting covert channels in DNS using character frequency analysis. Karasaridis [41] used the approach of histograms' calculations of request/response packet sizes using the fact that tracking and detecting the changes in the frequencies of non-conforming packets sizes lead to possible identification of DNS anomaly. Yuchi et al. [42] investigated DNS anomalies in the context of Heap's law stating that a corpus of text containing N words typically contains on the order of $cN^\beta$ distinct words, for constant c and $0<\beta<1$. Cermák et al. [43] identified that only four DNS packet fields are useful for most of the DNS traffic analyzing methods: queried domain name, queried record type, response return code and response itself. They use the concept of standard flow and extended flow to detect DNS attacks on large networks. The Statistical Feature String reflects statistical features of the domain name, similarly to ideas in [17-18], and includes the domain name entropy, an occurrence of bigrams within the domain name, the domain name zone and the domain name length and is used to calculate similarity distance between two different domain names during the clustering phase. The research described in [44] aimed detect the botnet traffic by inspecting the following parameters: Time-Based Features (Access ratio), DNS Answer-Based Features (Number of distinct IP addresses), TTL Value-Based Features, Domain Name-Based Features (% of numerical in domain name). In [45] the authors presented an approach in which the flow of DNS traffic between the source and the destination DNS server is used to detect attacks. They present Cross-Entropy Anomaly detection model which is used to detect anomalies in DNS packet sizes. For taking care of Feature-Based Detection, variations of entropy-based learning mechanisms were developed [46]. Based on the definition of context, there is a cause-effect relation among the features that characterize the context C and the corresponding consequences. In general, the features identifying context consequences are: numerical, descriptive and time/location-based features. We note that some past attempts were made in order to bring unsupervised machine learning mechanisms to deal with DNS related anomalies, see [47-49].

# 3 Methods

To define formally the problem of approximate string matching on temporal data for anomaly detection and to better understand the algorithm, we first present several basic definitions. These definitions will be used in the description of the methods.

## 3.1 Definitions

**Definition 1.** An *entity* $e \in E$ is an object that we observe to determine whether its behavior is normal (e.g. IP in the context of network traffic).

**Definition 2.** A *variable* $v \in V$ is an attribute name, which represents the entity $e$ (e.g. number of packets sent by some IP).

**Definition 3.** A *time-point tp* is defined by a quadruple (*entity*, *variable*, *timestamp*, *value*). The *value* is a numeric field that belongs to the *variable* at specific *timestamp*.

**Definition 4.** A *time-points series* ($TPS_{e,v}$) is a set of time-points with entity $e$ and variable $v$ sorted by lexicographical order. (Given two different sequences of the same length, $a_1a_2...a_k$ and $b_1b_2...b_k$, the first one is smaller than the second one for the lexicographical order, if $a_i<b_i$, for the first index $i$ where $a_i$ and $b_i$ differ.) Time-points series *suffix* of $TPS_{e,v}$ having length $t$ is denoted by $TPS^t_{e,v}$.

**Definition 5.** A *symbolic time interval*, $STI_{e,v} = <b, f, sym>$, is time-points series $TPS_{e,v}$ having timestamps that represent continuous time range, where $b$ represents the starting time-point timestamp and $f$ represents the last time-point timestamp. The *sym* is determined roughly by the time-points series gradient. A *symbolic time intervals series*, $STIS_{e,v}$ is a set of symbolic time intervals.

**Definition 6.** An *endpoint ep* from $STI_{e,v} = <b, f, sym>$, is defined by a quadruple $<timestamp, variable, sym, open>$, where the $STI_{e,v}$ is split to two endpoints, $ep_b$ and $ep_f$. The *open* is a Boolean field representing whether it is $ep_b$ or $ep_f$. The endpoint $ep_b$ equals $<b, v, sym, true>$ and $ep_f$ equals $<f, v, sym, false>$.

**Definition 7.** A *string $ds_e$* for entity $e$, is a list of endpoints sorted by lexicographical order. A *numerical string $s_e$* for entity $e$, is a list of numerical values, where each element is a transformation from $<variable, sym, open>$ fields to a unique numerical value. Let $|s_e|$ be the length of $s_e$.

**Definition 8.** A *pattern $P_x(s)$* contains the last $x$ elements of some string $s$ (i.e. $P_x(s)$ is the suffix of $s$ and $|P_x(s)| = x$). A *time-pattern $TP_x(ds)$* contains the last $x$ timestamp fields of elements from string $s$.

**Definition 9.** A *text $T(s)$* for some string $s$ is defined by $s \setminus P_x(s)$, i.e. $T(s)$ is the prefix of $s$, having length $n = |s| - x$.

**Definition 10.** We say that pattern $P_x(s)$ *($\alpha$, $\beta$)-occurs* with shift $t$ in text $T(s)$ if $|T(s)[t+j] - P_x(s)[j]| \leq \alpha$, $0 \leq j \leq x-1$ and $\sum_{j=1}^{n} |T(s)[t+j] - P_x(s)[j]| \leq \beta$. The parameter $\alpha$ controls the possible difference between each element in $P_x(s)$ versus the corresponding element in $T(s)$, while the parameter $\beta$ gives an upper bound on the total difference between all elements in $P_x(s)$ versus all elements in $T(s)$. This definition assumes that the elements of both pattern $P_x(s)$ and text $T(s)$ are represented by numerical values. In general, if we take $\alpha=\infty$, we will obtain a definition measuring only the distance between $P_x(s)$ and the corresponding subsequences in $T(s)$; the value of parameter $\alpha$ allows us to restrict the subsequences in $T(s)$ to have some specific form.

**Definition 11.** An *($\alpha$, $\beta$)-Approximate String Matching Problem of $P_x(s)$ in $T(s)$* is to determine whether a pattern $P_x(s)$ *($\alpha$, $\beta$)-occurs* with some shift in $T(s)$.

**Definition 12.** Anomalies are events in data that do not conform to a well-defined notion of normal behavior [27]. In our context we refer to Section III.F for anomaly identification.

**Definition 13. Approximate Matching on temporal data for Anomaly Detection:**

Given a set of entities $E$, where each entity $e \in E$ is described by time-points series $TPS_{e,v}$, the goal is to find, for each $e \in E$, a set of time ranges that represent anomalies.

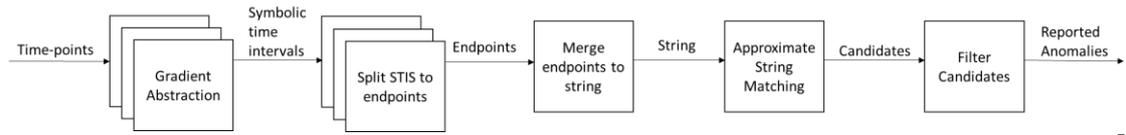

**Fig. 1.** Approximate Matching under Gradient Abstraction for anomaly detection components

### 3.2 Approximate Matching under Gradient Abstraction for anomaly detection Algorithm

The algorithm for Approximate Matching under Gradient Abstraction for anomaly detection, as it is shown in Algorithm 1, includes the following. The input for the algorithm is a list of time-points, each time-point *tp* is represented by (*entity*, *variable*, *timestamp*, *value*), parameter *x* that represents the duration of pattern $P_x(s)$ (i.e. $|P_x(s)| = x$) and $α, β$ parameters that control the possible difference between the elements in $P_x(s)$ versus the corresponding elements in $T(s)$. The output includes a list of anomalies with time-ranges associated to the entities in *E*. The main components of the algorithm, as shown in Fig. 1, are: (a) Gradient Abstraction which transforms the time-points series $TPS_{e,v}$ to symbolic time intervals series $STIS_{e,v}$, for each entity *e* and variable *v*; (b) Split operation that converts each symbolic time interval $STI_{e,v}$ to two endpoints $ep_b$ and $ep_f$ representing the start and finish time of $STI_{e,v}$ respectively; (c) Merge operation over the endpoints elements for entity *e* to a string $s_e$ sorted by lexicographical order; (d) Approximate string matching based on efficient data structure (in polylogarithmic time) in order to produce a list of candidates patterns; (e) Filtering of candidates to reduce false-positive results by removing some candidates using the pattern $P_x(s)$ and time-pattern $TPx(s)$. This algorithm runs online (every few minutes) to report recent anomalies.

The following sections present detailed explanation of each step described above.

### 3.3 Gradient abstraction

In the gradient abstraction phase, the time-points series $TPS_{e,v}$ are transformed into symbolic time intervals series $STIS_{e,v}$ for every entity *e* and variable *v*, according to the gradient abstraction technique. The main idea is to assign the symbols for the symbolic time intervals based on the time-points series gradient. The symbols could be Increasing (*I*), Decreasing (*D*) and Stable (*S*) according to the time-points series trend, similarly to [11]. For better representation, *I* and *D* symbols are divided to three sub-symbols: High (*H*), Medium (*M*) and Low (*L*), according to the trend intensity. An illustration is shown in Fig. 3.

| **Algorithm 1 – Approximate Matching under Gradient Abstraction for Anomaly Detection** | |
|---|---|
| **Input:** | time-points *tps* |
| **Input:** | pattern length *x* |
| **Input:** | approximation bounds $α, β$ |
| **Output:** | list of reported anomalies with time ranges |
| 1: | *found_list* = [] |
| 2: | *cold_start_list* = [] |
| 3: | **for** *e* **in** *E*: |
| 4: |   *list_of_eps* = [] |
| 5: |   **for** *v* **in** *V*: |
| 6: |     *STIS* = gradient_abstraction(*tps*) |
| 7: |     *list_of_eps*.append(split_STIS (*STIS*)) |
| 8: |   **end for** |
| 9: |   $ds_e$ = merge_eps_to_string(*list_of_eps*) |
| 10: |   $s_e$ = transform_to_numeric($ds_e$) |
| 11: |   $T(s) = s_e [0: |s_e| - x]$ |
| 12: |   $P_x(s) = s_e [|s_e| - x : |s_e| -1]$ |
| 13: |   $TP_x(ds) = ds_e [|ds_e| - x : |ds_e| -1]$ |
| 14: |   *is_found* = approx_string_matching($T(s), P_x(s), α, β$) |
| 15: |   **if** (*is_found*): |
| 16: |     *found_list*.append($P_x(s), TP_x(ds)$) |
| 17: |   **else**: |
| 18: |     *cold_start_list*.append($P_x(s), TP_x(ds)$) |
| 19: | **end for** |
| 20: | *reported_anomalies*=filter(*found_list*, *cold_start_list*) |
| 21: | **return** *reported_anomalies* |

The gradient abstraction phase is done by comparing the current time-point value to a previous time-points series values and not only to the last time point value, in order to avoid being influenced by noisy data. For example, if we get time-points series values of (200, 190, 180, 170, 200) representing the number of packets in a minute for some particular IP, by considering only the last element value, we may think that there is an Increasing trend (*I*) from 170 to 200, but when we are looking at the entire series, we probably will get to conclusion that the trend is Stable (*S*).

For gradient abstraction, we use the following *angle*, *slope* and *relation* definitions to create symbolic time intervals.

**Definition 14.** The *angle*($p_1$, $p_2$) in the plane between the two points $p_1 = (x_1, y_1)$ and $p_2 = (x_1+1, y_2)$, is defined by the angle between the line passing through these points and the axis X. In other words,

$$\text{angle}(p_1, p_2) = \cos^{-1}\left(\frac{1}{\sqrt{(y_2-y_1)^2+1}}\right) \quad (1)$$

**Definition 15.** A *slope*($TPS_{e,v}$, $t$) is defined by the average value of angles between pairs of the adjacent time-points series values of the last *t* time-points in $TPS_{e,v}$. See pseudo-code in Fig. 2.

**Definition 16.** A *relation*($TPS_{e,v}$, $t$) is defined by the current time-point value divided by the average value of the last *t* time-points values in $TPS_{e,v}$. See pseudo-code in Fig. 2.

```
slope(TPS_{e,v}, t):
    for tps_1.value, tps_2.value in TPS_{e,v}:
        p_1 = (0, tps_1.value), p_2 = (1, tps_2.value)
        if tps_1.value > tps_2.value:
            angle(p_1, p_2) = -angle(p_1, p_2)
        sum_angle += angle(p_1, p_2)
    slope = abs(sum_angle / t-1))
    return slope

relation(TPS_{e,v}, t, cur_value):
    avg_prev_tps_suffix_values = ∑(TPS_{e,v}.value) / t
    relation = cur_value / avg_prev_tps_suffix_values
    return relation
```

**Fig. 2.** Slope and relation calculation pseudo-code

Using *slope* and *relation*, we define seven symbols as it is shown in Table 1. We consider the *relation* value in addition to the *slope* value, since high *slope* values alone not necessarily indicate significant change in data trend. For example, the increase in data volume from 1 to 2 leads to 45° (relatively high value) of *slope* value without providing meaningful insight. In the evaluation, we set the thresholds for *slope* and *relation* values. For *relation* that equals to 1, we set the symbol as *S*.

**Table 1.** Symbols creation according to slope and relation values

| Slope  | Relation              | Symbol      |
|--------|-----------------------|-------------|
| High   | High (1 / High)       | I-H (D-H)   |
| High   | Medium (1 / Medium)   | I-M (D-M)   |
| High   | Low (1 / Low)         | I-L (D-L)   |
| Medium | High (1 / High)       | I-M (D-M)   |
| Medium | Medium (1 / Medium)   | I-M (D-M)   |
| Medium | Low (1 / Low)         | I-L (D-L)   |
| Low    | High (1 / High)       | S (S)       |
| Low    | Medium (1 / Medium)   | S (S)       |
| Low    | Low (1 / Low)         | S (S)       |

As we mentioned earlier, the gradient abstraction technique has a few advantages in our case. First, to find an anomaly, we are looking for a change, (e.g. increasing volume of traffic) while stable trend is less relevant. Second, we significantly reduce the amount of data per each entity $e$, since instead of many time-points series $TPS_{e,v}$, we look only on the change in data. Most entities behave normally and the original time-points series contain thousands of samples for a period, while with gradient abstraction we can represent this period of time with a small series of symbolic time intervals $STIS_{e,v}$, enabling to find anomalies very fast.

### 3.4 Split STI to endpoints

In the split phase, each symbolic time interval $STI_{e,v}$ is converted into two endpoints $ep_b$ and $ep_f$, where $b$ represents the begin time and $f$ represents the finish time of $STI_{e,v}$. We convert the symbolic time interval to two time-points, since we are interested in the periods, in which the change in the data starts and when it ends. This technique allows us to simply work with ordered data rather than work with time interval symbols and their relations.

### 3.5 Merge endpoints to string

In the merge phase, all the endpoints for entity $e$ and each $v \in V$ are sorted by a lexicographical order, to produce a string $ds_e$. The motivation to work with a string is coming from the fact that we are interested in finding occurrences of continuous patterns in the resulted string. Moreover, using a string representation allows us to perform approximate string matching as it is shown in the next section.

### 3.6 Approximate String Matching

In $(α, β)$-Approximate String Matching, we check if pattern $P_x(s)$ $(α, β)$-occurs with shift $t$ in text $T(s)$. If we found a match we insert the pattern to the *found_list* of *potential anomalies*, otherwise we insert it to the *cold_start_list*. The elements of $P_x(s)$ and $T(s)$ are represented by numerical values in the following way. Note that an endpoint is defined as *<timestamp, variable, sym, open>*, in which the last triple fields should be transformed. We define a numeric representation using two digits for *variable* and *sym* fields that allow us to handle 100 distinct values between '00' to '99' for each field and one digit for the *open* field. The overall numerical element is defined using five digits: the first digit is for the *open* field, the following two digits for the *sym* field and the last two digits for the *variable* field. For the *sym* field we convert the symbols *<D-H, D-M, D-L, S, I-L, I-M, I-H>* to *<00, 01, 02, 03, 04, 05, 06>*, respectively to represent close trends by close values. For the *variable* field, related variables are represented by close values according to the specific domain and the *open* field is converted from *true* and *false* to 2 and 1, respectively. The subtract operation, $|T(s)[t+j]-P_x(s)[j]|$ is a regular subtraction of two numeric values, e.g. $v_1$-$I$-$H_b$ − $v_2$-$I$-$M_b$ equals to 20600 − 20501 = 99. For $\alpha \geq 99$ these two elements are matched. The parameter $\beta$ is defined according to the $L_1$ or $L_2$ metrics (as described later) and upper bounded by $x \cdot \alpha$. In the next section, the approximate string matching analysis is shown in detail.

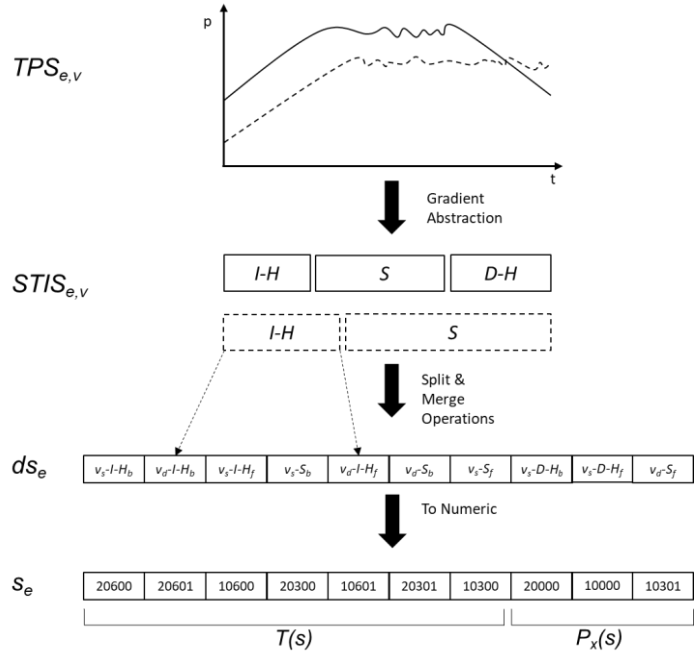

**Fig. 3.** Illustration of the flow from time-points series $TPS_{e,v}$ to a string $s_e$

Fig. 3 illustrates the components described above for some entity $e$. In the upper part, there are two lines of time-points series $TPS_{e,v}$ for two variables $v_s$ and $v_d$, representing the volume of data over time. In the middle part, the gradient abstraction phase transforms time-points belonging to $TPS_{e,v}$ to symbolic time intervals series $STIS_{e,v}$. The symbols inside of solid line boxes belong to the variable $v_s$ and the symbols inside the dashed line boxes come from variable $v_d$. At the bottom part, the split and merge phases are shown, where we first split each $STI_{e,v}$ to two endpoints (e.g. *I-H* symbol in solid line is split to $v_s$-$I$-$H_b$ and $v_s$-$I$-$H_f$) and then merge all endpoints by lexicographical order to produce an ordered string $ds_e$ for each entity $e$. Finally, we transform $ds_e$ to numeric string $s_e$ and extract $P_x(s)$ and $T(s)$ to perform string matching.

### 3.7 Filter candidates

In the filter candidates' phase, there are two input lists, the *found_list* containing the detected potential patterns and the *cold_start_list* containing mismatched patterns that may represent possible anomaly occurrence. In order to remove irrelevant candidates from further consideration, we perform the following actions: First, we seek for an increase or decrease in the data trend and remove pattern candidates that do not contain numerical values corresponding to either *I-M*, *I-H*, *D-M* and *D-H*. For the cold start candidates, we keep only patterns with significant change such as *I-H* and *D-H*. Second, candidates with a long duration are less likely to be an anomaly (and even if it is indeed such, the suspected anomaly duration is too long for effective investigation) and therefore can be removed.

## 4 Approximate String Matching Analysis

In order to identify a potential string matching, we can follow exact and approximate solutions.

### 4.1 Exact matching

There are several approaches that can allow us to perform string matching, however they work only under the assumption of exact matching which leads to significant deterioration in our ability to identify anomalies. The first and the simplest approach is to consider all possible continuous patterns in text $T(s)$ (up to the length $x$ of $P_x(s)$) and to build a hash table containing all of them, where each pattern in hash table is associated with the value representing the number of times this pattern appears in text. Overall the construction expected time is $O(nx^2)$ and query expected time is $O(x)$. The second approach is to use suffix tree, see [50]. This will allow us to preprocess the text in $O(n)$ time and obtain query time $O(x)$. If we want to count all of the appearances of $P_x(s)$, the query time grows up to $O(x + z)$, where z is the number of such appearances. We can also generalize the first approach to find the lower and upper bound on the number of appearances of $P_x(s)$ by considering, instead of all possible continuous patterns in $T(s)$, only those having length $i = 1,2,...,2^{\log x}$. Then, we can compute the upper and the lower bounds on the number of appearances using segment tree structure [51] in $O(\log x)$ query time after preprocessing time $O(nx + n \log n)$.

### 4.2 Data structure for approximate pattern matching

Here we propose an efficient data structure in order to perform approximate pattern matching queries. Of course, we can use a recent result obtained in [52] that solves the static version of the problem in linear time, but the goal is to obtain better bounds.

There are also three related results, although not exactly comparable with our goal. The paper by Cole et al. [53] presents result for approximate pattern matching under Hamming distance, up to k errors, and the query time is $O(\log^k n)$ working only with static text. Dynamic text (changes in front and back) was done with a changing pattern, in other words, one has a given pattern and you can make changes in it. The polylogarithmic update and query time were described in [54]. Here, no approximate matching is considered. Finally, one can consider using some hashing technique, for example, locality-sensitive hashing LSH (see [55]) or Learning to Hash methodology [56]. LSH hashes input items so that similar items map to the same "buckets" with high probability (the number of buckets being much smaller than the universe of possible input items). In this case, we indeed can perform an approximate matching search, but the algorithm is randomized, i.e. it succeeds in finding a match within distance $O(R)$ from $P_x(s)$ (if there exists a match within distance $R$) with high probability. The preprocessing time is roughly $O(n^{1+\epsilon}x)$ and query time is $O(n^\epsilon x)$, for any given $\epsilon > 0$. Moreover, in order to count all such matches (this is supported by our technique described later), the algorithm may require a linear time. Regarding learning to hash methodology – it's a data dependent hashing approach which aims to learn hash functions from a specific dataset so that the nearest neighbor search result in the hash coding space is as close as possible to the search result in the original space, and the search cost as well as the space cost are also small, see for example [57]. However, the all known results (see survey [56]) have only empirical performance bounds.

Our data structure is based on efficient use of balanced binary search trees together with geometric observations that allow us to perform updates and queries efficiently. Let the length of $P_x(s)$ be $x$, and the length of $T(s)$ is $n$ (without inclusion of $P_x(s)$). We consider the pattern $P_x(s)$ as the point in $x$-dimensional space, where the $i$-th measurement in $P_x(s)$ corresponds to the $i$-th coordinate of the corresponding point. The text $T(s)$ can be viewed as a collection of $n$-$x$+$1$ points in $x$-dimensional space, obtained be considering $x$ consecutive measurements of $T(s)$ starting from the beginning and sliding them each time by one position to the right. What does it mean geometrically that there is an approximate match between $P_x(s)$ represented by the point in $x$-dimensional space and some of the $n$-$x$+$1$ points obtained from text $T(s)$, from the point of view of the parameter $\alpha$? It means that the distance from each coordinate value of $P_x(s)$ point can be at most $\alpha$ to the corresponding coordinate of value of other point. In other words, all the points that are located inside of $x$-dimensional cube centered at point $P_x(s)$ and having side length of $2\alpha$ represent an approximate match for $P_x(s)$ in terms of parameter $\alpha$. See Fig. 4 below for 2-dimensional case.

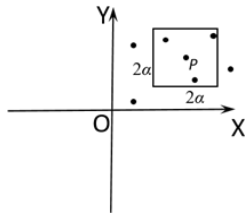

**Fig. 4.** Query point $P_x(s)$ and all the points inside of the square are at distance $\alpha$

This can be done by adopting Chazelle's [58-59] orthogonal range counting approach. Namely, given a set of $n$ points in the plane and an orthogonal range, we want to find the number of points contained in the orthogonal range. Chazelle proposes a data structure that can be constructed in time $O(n \log^{x-1} n)$ and occupies $O(n)$ space, such that a range-counting answer for a query region can be answered in time $O(\log^{x-1} n)$ with an update time of $O(\log^x n)$. The major idea is to keep $x$-dimensional balanced binary tree and to (sequentially) perform a number of binary searches in each one of the trees corresponding to separate dimensions by maintain the obtained filtered results in the canonical sets. The last level tree binary searches can be avoided by using fractional cascading technique, see [59]. However, this technique does not allow dynamic updates of the structure. In order to overcome the difficulty, an additional multiplicative factor of $O(\log \log n)$ time should be added to the update operations [60].

So far, we have dealt only with the parameter $\alpha$, but what happens with our second requirement that depends on the parameter $\beta$? Here it looks that the problem is more complicated since we need to consider the sum of the distances between the point $P_x(s)$ and all other points, and to bound this sum by $\beta$ value. Fortunately, we observe the following.

**Observation 1.** The sum of the distances between each coordinate of point $P_x(s)$ and any other point in $x$-dimensional space is actually the $L_1$ distance between them.

This observation, in fact, provides us an efficient way to count all the points in $x$-dimensional space that satisfy the limitation of $\beta$ parameter. In this case, the query becomes the titled by 45° $x$-dimensional cube, which is centered at $P_x(s)$, having the size length of $\beta$. See Fig. 5 for 2-dimensional case. We notice that this query is performed on the canonical subsets (their number is at most $O(\log^{x-1} n)$) obtained after the step 1 for an appropriate selection according to the parameter $\alpha$.

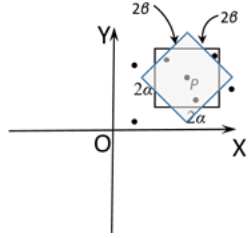

**Fig. 5.** Query by titled shadowed square in order to satisfy constraint $\beta$

Here we face the problem that in order to make the query by the titled square centered at $P_x(s)$ using the same approach as before, we need our points to be sorted according to the rotated axes; otherwise we cannot perform the standard binary search. We are not able to afford the rotation of axes since it may lead to a linear time query time, while our goal is to have polylogarithmic complexity. In order to understand how we can overcome this difficulty, we have to explain the specific construction of the multi-level balanced binary search tree that we use in the first step of the solution while searching the points having distance of at most $\alpha$ from one of the $P_x(s)$ coordinates. A range tree on a set of 1-dimensional points is a balanced binary search tree on those points. The points stored in the tree are stored in the leaves of the tree; each internal node stores the largest value contained in its left subtree. A range tree on a set of points in $x$-dimensions is a recursively defined multi-level binary search tree. Each level of the data structure is a binary search tree on one of the $x$-dimensions. The first level is a binary search tree on the first of the $x$-coordinates. Each vertex $v$ of this tree contains an associated structure that is a $(x-1)$-dimensional range tree on the last $(x-1)$-coordinates of the points stored in the subtree of $v$. Here comes the crux: in our case, each vertex $v$ of the last level of our tree will contain an associated structure of $x$-dimensional range tree in the rotated system of axes. In this way, when we finish with the parameter $\alpha$ query having $O(\log^{x-1} n)$ canonical subsets, we will continue to work on parameter $\beta$ having the same subsets but in rotated system of axes. Eventually, the number of subsets may become to be $O(\log^{2x-1} n)$ which will dominate the query time. The update time becomes $O(\log^{2x} n)$. Here we want to mention that we did not made an intensive attempt to reduce further the update and the query time of our operation. Our goal was to show that it is possible to perform approximate pattern matching queries in polylogarithmic time with similar update times. One possible direction to improve the time can be by dividing the titled square in to the number of wedges as follows: Let $l_1$ be the line parallel to the $x$ axis and passing through $P_x(s)$, $l_2$ be a line whose slope is 45° passing through $P_x(s)$, $l_3$ be the line parallel to the $y$ axis and passing through $P_x(s)$ and $l_4$ a line whose slope is 135° passing through $P_x(s)$. These lines define wedges (1) $Q_1$ – the wedge of points between $l_1$ and $l_2$ whose $x$ coordinates are larger than their $y$ coordinates, (2) $Q_2$ – the wedge of points between $l_2$ and $l_3$ whose $y$ coordinates are larger than their $x$ coordinates, and (3) $Q_3$ – the wedge of points between $l_3$ and $l_4$ whose $y$ coordinates are larger than their $x$ coordinates. We can build in such fashion all 8 wedges around the point $P_x(s)$. Now, instead of looking for the sum of distances (as parameter $\beta$ dictates), we can look, per each wedge, only for dominating coordinate that might be easier and may lead to the speedup in the solution. See [61] for the use of this technique.

Note that while the parameter $\alpha$ controls the possible absolute difference between each measurement in $P_x(s)$ versus corresponding measurement in $T(s)$, it can have a different interpretation. For example, it may measure the squared distance between each measurement in $P_x(s)$ versus corresponding measurement in $T(s)$. The same holds for the parameter $\beta$ that can provide an upper bound on, e.g. the rooted sum of squared differences between corresponding measurements of $P_x(s)$ and $T(s)$. In other words, instead of computing the standard $L_1$ $\beta$ threshold, we can have a new $L_2$ $\beta$ threshold:

```
L₁β-threshold(Pₓ(s) , T(s)):
    total = 0
    for i in range(x):
        total += abs(Pₓ(s)[i] - T(s)[i])
```

```
L₂β-threshold(Pₓ(s) , T(s)):
    total = 0
    for i in range(x):
        total += (Pₓ(s)[i] - T(s)[i])²
    total = √total
```

In our geometric interpretation, this case of *β* will mean that the titled square shown in Fig. 5 will become a disk centered at *Pₓ(s)*, having a diameter of *2β*.

Here, we would like to emphasize how our *theoretical* result relates to the HOT SAX [20-21] approach which looks for the "most unusual subsequence" in the given sequence and have worst-case quadratic running time. As we have shown, under L₁ metric our solution has a subquadratic time and, in fact, works for any value of *α*, opposite to [20-21] assuming *α*=∞. But HOT SAX tries to compare subsequences under L₂ metric which makes the problem harder. Still, using techniques from computational geometry, we can find the solution in subquadratic time, for any fixed, *x*-dimensional space. Below we describe a number of possible approaches to do that. First, we observe that our problem is equivalent to the following problem: given a set of *n* points in the *x*-dimensional space, we want to report one of given points having the Euclidean distance to its nearest neighbor point maximized over all possible choices. We can deal with this problem as follows:

1) First approach

The papers [62-63] present solutions to the dynamic, bichromatic closest-pair problem in the plane. In the bichromatic closest-pair problem we are given two sets of points *S* and *T* and we are required to compute the closest pair of points $(u, v)$ such that $u \in S$ and $v \in T$. The dynamic data structure, allowing insertions and deletions of points, presented in [62-63] is used to find, given a point *p*, a point *q* $\in T$ minimizing the Euclidean distance between *p* and all other points. Doing this for all *n* choices of *p* and noticing that each update and query time is $O(n^{1/3+\varepsilon})$, we obtain a total subquadratic runtime of $O(n^{4/3+\varepsilon})$ for the planar case. This approach can be generalized to *x*-dimensional space using the techniques of [65,70] but the time bounds for exact nearest neighbors have the form $O(n^{1-\varepsilon(x)})$ for constant $\varepsilon(x)$ that get very small as *x* gets large.

2) Second approach

In a very recent paper [66] (see also [67]), Chan presents a data structure for the planar point set that can be preprocessed in $O(n \log n)$ time, having $O(\log^2 n)$ amortized insertion time, and $O(\log^4 n)$ amortized deletion time, so that we can find the nearest neighbor to any query point in $O(\log^2 n)$ time. It improves the first approach described above and has a total $O(n \log^2 n)$ runtime.

3) Third approach

Efrat et al. [68] presented a dynamic soft nearest-neighbor data structure that maintains a dynamic set of points in *x*-dimensional space, *S*, subject to insertions, deletions, and soft nearest-neighbor queries (all in $O(\log n)$ time): given a query point *q*, return either of the following: *the nearest neighbor p\* of q* in *P* or a pair of points in *P* having distance between them less than the distance between *q* and *p\**. Having this data structure, we can find the overall solution in *O(n* log *n)* time by identifying the point with the largest distance to its nearest neighbor.

4) Fourth approach

We can take an advantage of the fact that we need to compute the nearest neighbor to each of the given points. Thus, we can avoid doing queries and try to compute the entire solution without involving complicated data structures. The papers [64,69] solve the following problem: Given a fixed dimension, a semi-definite positive norm (thereby including every L$_p$ norm), and *n* points in this space, the nearest neighbour of every point can be found in $O(n \log n)$ time and the *m* nearest neighbours of every point can be found in $O(mn \log n)$ time. In our particular case, for a fixed *x*-dimensional space, we can have a solution in $O(n \log n)$ time.

All these described solutions are exact; there are many geometric solutions that find approximate nearest neighbors in high dimensional space, see e.g. [71], that can be used if we are willing to speed-up the runtimes versus the accuracy we obtain.

## 5 Experimental Setup

### 5.1 Dataset

We have evaluated our method on the DARPA 2009 dataset from IMPACT Cyber database. The DARPA 2009 dataset is created with synthesized traffic to emulate traffic between 172.28.0.0/16 and the Internet. This dataset has been captured in 10 days between the 3rd and the 12th of November of the year 2009. It contains synthetic DNS background data traffic. The dataset has a variety of security events and attack types that describes the modern style of attacks. This dataset has been already evaluated using the supervised learning techniques, see for example [72].

We have evaluated the raw traffic for a few IPs to see the anomalies. Fig. 6 shows the sum of packets for IP 172.28.10.6 which serves as the Firewall. The traffic trend is periodical, where from 14:00PM to 12:00AM there are ~200K packets per hour and from 12:00AM to 14:00PM there are ~75K packets per hour. At the 3-Nov and at the 12-Nov there are two peaks in traffic, which are reported as DNS attacks.

### 5.2 Baseline algorithms

We compared our approach with the followings anomaly detection over time baselines, that were implemented to run them on this dataset:

1. Approximate String Matching (ASM) [52]—this approach is based on time-points, which means considering the real values summarized by unit time periods instead of trends.
2. Auto-regression (AR) model [73] based on time-points—the model learns the best lag based on statistics tests.
3. Linear regression (LR) based on time-points [74].
4. Lasso based on time-points [75].
5. Random Forest regressor (RF) based on time-points [76].
6. K-Nearest Neighbors (KNN) regressor based on time-points [77].

### 5.3 Experimental setting

The variables we have looked at are:

**Variable A**: The total number of DNS packets per minute in the traffic.

**Variable B**: The number of transmitted DNS packets per minute, per each IP.

The evaluated dataset has been provided with the ground truth events by the dataset creators. If the anomaly has been detected by our method within time interval that overlaps with the corresponding time interval in ground truth events file, we report this as True Positive (TP). If there is no such overlapping interval in ground truth events file, we report this as False Positive (FP). If there is any interval in ground truth events file which was not hit by any of our identified intervals of anomalies, it is considered as False Negative (FN). All other cases had been treated as true negatives. Since we are looking for anomalies, we notice that the True Negative (TN) is a huge number and, thus, the classical metrics such

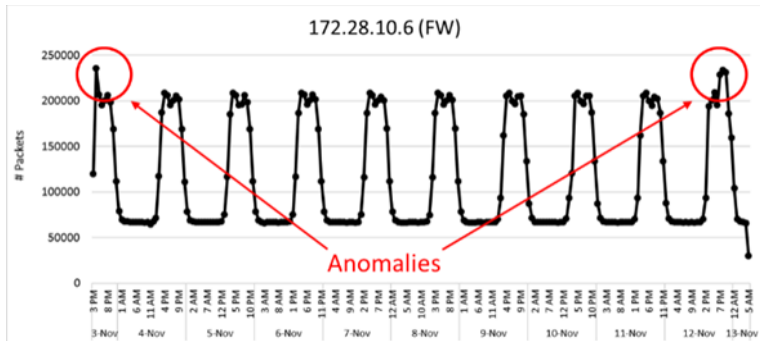

**Fig. 6.** Traffic of specific IP with 2 anomalies

as accuracy, AUC, etc., are not well suited for this task

We fixed the parameters for the *slope* and *relation* for the entire evaluation: *high_slope_thresh* = 45°, *low_slope_thresh* = 15°, *high_relation_thresh* = 2, *low_relation_thresh* = 1.5.

## 6 Results

### 6.1 Experiment 1 – THAAD vs. Anomaly detection baselines

We compare our proposed method to the baselines as shown in Table 2 by also studying True Positive and False Negative rates (TPR/FNR). As we can learn from the table, our approach outperformed the others, with the next best method being ASM, which uses similar strategy based on time-points. Since THAAD and ASM are the only methods that use approximate matching, we run THAAD with exact matching ($\alpha=0$), which we can see that it is better from other exact methods.

Table 2. Comparison of THAAD to baselines

| Method | FP | FN | TPR | FNR |
|---|---|---|---|---|
| **THAAD** | **3** | **4** | **0.991** | **0.008** |
| ASM | 3 | 12 | 0.974 | 0.025 |
| Exact THAAD | 1 | 19 | 0.958 | 0.041 |
| AR | 1 | 30 | 0.935 | 0.064 |
| KNN | 1 | 35 | 0.924 | 0.075 |
| LR | 1 | 53 | 0.885 | 0.114 |
| LASSO | 1 | 53 | 0.885 | 0.114 |

### 6.2 Experiment 2 – approximate matching analysis

We define several accumulated levels for the parameter $\alpha$:

1. $\alpha=0$, for exact matching.
2. $\alpha=1$, for close variables substitution. We allow matching between A and B variables with the same symbol.
3. $\alpha=100$, for close trend. We allow matching between High to Medium, Medium to Low and Low to Stable.
4. $\alpha=750$, for unrestricted $\alpha$, i.e. only $\beta$ parameter is applied.

Obviously, if $\alpha$ value goes up, flexible matching between the symbols trends is allowed. We ran this experiment using parameter $\beta$ as $L_1$ and $L_2$ metrics. The obtained results were identical for both metrics.

The approximation was evaluated by $\alpha$ and $\beta$ parameters using the entire dataset as shown in Table 3 and Fig. 7. If $\alpha$ and $\beta$ values grow up, the False Positive results are increased and the False Negative results are decreased, since more patterns are reported as anomalies (some of them are real and some are not). The best result was obtained with $\alpha=100$, having $\beta>=200$, which means that both feature substitution and close trend are allowed.

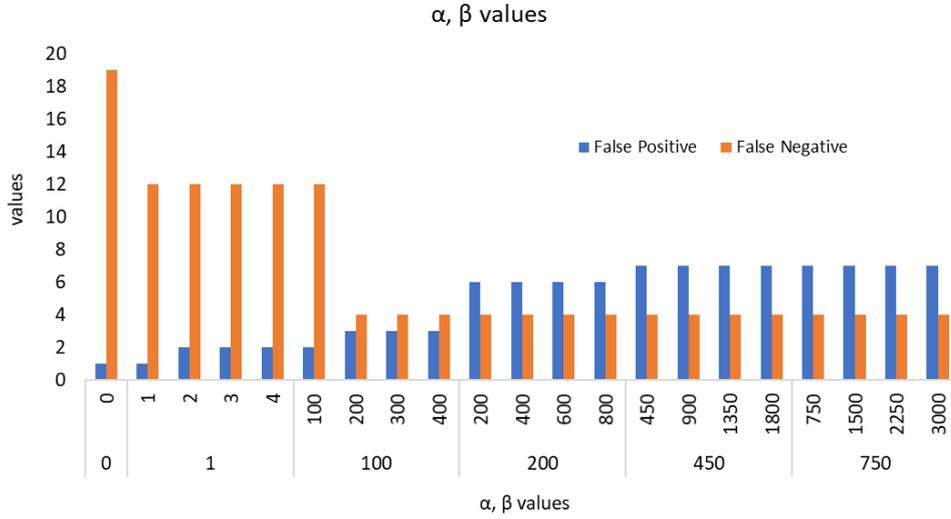

**Fig. 7.** Performances for α, β values. α={0,1,100,200,450,750} and β ={1·α,2·α,3·α,4·α}

**Table 3.** Approximate Matching

| α | B | TP | FP | FN | TPR | FNR | F1 |
|---|---|---|---|---|---|---|---|
| 0 | 0 | 444 | 1 | 19 | 0.959 | 0.041 | 0.978 |
| 1 | 1 | 451 | 1 | 12 | 0.974 | 0.026 | 0.986 |
| 1 | 2 | 451 | 2 | 12 | 0.974 | 0.026 | 0.985 |
| 1 | 3 | 451 | 2 | 12 | 0.974 | 0.026 | 0.985 |
| 1 | 4 | 451 | 2 | 12 | 0.974 | 0.026 | 0.985 |
| 100 | 100 | 451 | 2 | 12 | 0.974 | 0.026 | 0.985 |
| **100** | **200** | **459** | **3** | **4** | **0.991** | **0.009** | **0.992** |
| **100** | **300** | **459** | **3** | **4** | **0.991** | **0.009** | **0.992** |
| **100** | **400** | **459** | **3** | **4** | **0.991** | **0.009** | **0.992** |
| 200 | 200 | 459 | 6 | 4 | 0.991 | 0.009 | 0.989 |
| 200 | 400 | 459 | 6 | 4 | 0.991 | 0.009 | 0.989 |
| 200 | 600 | 459 | 6 | 4 | 0.991 | 0.009 | 0.989 |
| 200 | 800 | 459 | 6 | 4 | 0.991 | 0.009 | 0.989 |
| 450 | 450 | 459 | 7 | 4 | 0.991 | 0.009 | 0.988 |
| 450 | 900 | 459 | 7 | 4 | 0.991 | 0.009 | 0.988 |
| 450 | 1350 | 459 | 7 | 4 | 0.991 | 0.009 | 0.988 |
| 450 | 1800 | 459 | 7 | 4 | 0.991 | 0.009 | 0.988 |
| 750 | 750 | 459 | 7 | 4 | 0.991 | 0.009 | 0.988 |
| 750 | 1500 | 459 | 7 | 4 | 0.991 | 0.009 | 0.988 |
| 750 | 2250 | 459 | 7 | 4 | 0.991 | 0.009 | 0.988 |
| 750 | 3000 | 459 | 7 | 4 | 0.991 | 0.009 | 0.988 |

For unrestricted α value (i.e. α=750), only β was applied to bound the subsequences matching. As it shown in Fig. 8a, the best performance is achieved using β in the range from 90 to 300 with FP=6 and FN=4. For the same β values with restricted α values, the performance is better (e.g. for α=100, β =300 the results are FP=3 and FN=4 as shown in Table 3). Moreover, from Fig. 8b we can learn that for a fixed value of β=600, the value of α has a large impact on the performance. In this case, the equilibrium (between FP and FN) is by using α=100. It supports our claim that the parameter α has a significant influence on the precision of the results.

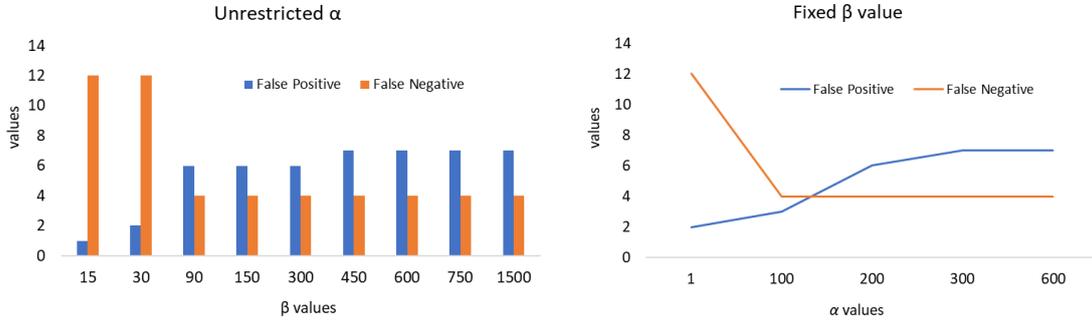

**Fig. 8.** (a) Performances for unresticted α=750. (b) Performances for β=600

Additionally, we evaluated the length $x$ of $P_x(s)$, which means how many symbols are in $P_x(s)$. The results depending on pattern length ($x$) are shown in Fig. 9. This experiment ran with $α=100$ and $β=300$. Length 5, performed the best with 3 and 4 for FP and FN results, respectively.

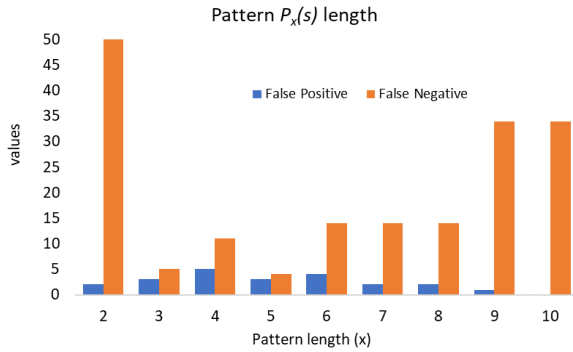

**Fig. 9.** Pattern length $x$ ($α=100$, $β=300$)

## 7 CONCLUSIONS

In this paper we had introduced the idea of identifying unusual subsequences efficiently using a new distance metric an addition to the Euclidean distance ($β$). The overall process starts with gradient abstraction to represent multivariate time series in a more abstract and compact fashion. After the data is transformed into a symbolic time intervals representation, we transform it into a sequence-based representation of the resulted time intervals' end points (begin and finish times). Then we identify unusual subsequences in the obtained sequence using effective algorithms and data structure based on geometric observations that supports polylogarithmic update and query times. We have introduced a new parameter ($α$) to control the pairwise difference between the symbols that belong to subsequences in addition to the standard distance metric – this allowed us to obtain more precise results in the context of anomalies. We evaluated the contribution of the additional parameter $α$ and found that the False Positive was improved, since $α$ filters some false patterns that can pass $β$ parameter when using alone.

By product, the new data structure produces a theoretically better worst-case runtime solution to the HOT SAX approach, breaking the quadratic worst-case runtime barrier. It would be interesting to run the HOT SAX solution combined with our data structure to understand the practical importance of improved worst-case running time scenarios. By using gradient abstraction in combination with approximate string matching we have shown simulatively that the detection of anomaly events in DNS data traffic can be done very accurately and efficiently, compared with previously known approaches.

The future steps for improvements will include: consideration of $α$ and $β$ parameters in the context of different distance measures (e.g. [81]) and identification of additional hidden trends in the historical data, e.g. by considering the historical data in reverse order.

16